\documentclass[sigconf,screen=true,bookmarks=false]{acmart}

\usepackage[utf8]{inputenc} %
\usepackage[T1]{fontenc}    %
\usepackage{xcolor}         %
\usepackage{booktabs}
\usepackage{algorithm}
\usepackage{graphicx}
\usepackage{threeparttable}
\usepackage{multirow}
\usepackage{amsmath}
\usepackage{bm}
\usepackage{bbm}
\usepackage{amsfonts}
\usepackage{amsthm}
\usepackage[noend]{algpseudocode}
\usepackage{enumitem} %
\usepackage[subrefformat=parens,labelformat=parens]{subfig}

\usepackage{wrapfig}

\definecolor{citecolor}{RGB}{34,139,34}
\definecolor{mydarkblue}{rgb}{0,0.08,1}
\definecolor{mydarkgreen}{rgb}{0.02,0.6,0.02}
\definecolor{mydarkred}{rgb}{0.8,0.02,0.02}
\definecolor{mydarkorange}{rgb}{0.40,0.2,0.02}
\definecolor{mypurple}{RGB}{111,0,255}
\definecolor{myred}{rgb}{1.0,0.0,0.0}
\definecolor{mygold}{rgb}{0.75,0.6,0.12}
\definecolor{myblue}{rgb}{0,0.2,0.8}
\definecolor{mydarkgray}{rgb}{0.,0.2,0.2}

\definecolor{lightred}{RGB}{255,235,235}
\definecolor{lightgreen}{RGB}{235,255,235}
\definecolor{lightblue}{RGB}{235,235,255}
\definecolor{lightcyan}{RGB}{235,255,255}
\definecolor{lightmagenta}{RGB}{255,235,255}
\definecolor{lightyellow}{RGB}{255,255,235}

\definecolor{qxkcolor}{RGB}{215,235,255}
\definecolor{softmaxcolor}{RGB}{230,235,255}
\definecolor{probxvcolor}{RGB}{255,255,235}

\definecolor{topkcolor}{RGB}{255,235,235}
\definecolor{zecolor}{RGB}{255,255,235}
\definecolor{dynacolor}{RGB}{235,255,255}

\definecolor{reviewcolor}{RGB}{0,0,200}

\newcommand{\calP}{\mathcal{P}}

\newcommand{\calS}{\mathcal{S}}
\newcommand{\calT}{\mathcal{T}}

\theoremstyle{plain}

\theoremstyle{definition}

\algdef{SE}[DOWHILE]{Do}{doWhile}{\algorithmicdo}[1]{\algorithmicwhile\ #1}%

\newcommand{\squishlist}{
 \begin{list}{$\bullet$}
  { \setlength{\itemsep}{0pt}
     \setlength{\parsep}{3pt}
     \setlength{\topsep}{3pt}
     \setlength{\partopsep}{0pt}
     \setlength{\leftmargin}{1.5em}
     \setlength{\labelwidth}{1em}
     \setlength{\labelsep}{0.5em} } }

\newcommand{\squishend}{
  \end{list}  }

\newcommand{\name}{\texttt{ADEPT-Z}\xspace}

\graphicspath{{./figs/}}
\usepackage{geometry}
\AtBeginDocument{%
  }

\setcopyright{acmlicensed}
\copyrightyear{2018}
\acmYear{2018}
\acmDOI{XXXXXXX.XXXXXXX}

\acmConference[ASP-DAC 2025]{Make sure to enter the correct
  conference title from your rights confirmation emai}{Jan. 20--23,
  2025}{Tokyo, Japan}
\acmISBN{978-1-4503-XXXX-X/18/06}
\usepackage{pifont}
\usepackage[normalem]{ulem}
\usepackage{xspace}
\algdef{SE}[DOWHILE]{Do}{doWhile}{\algorithmicdo}[1]{\algorithmicwhile\ #1}%

\begin{document}
\settopmatter{printacmref=false}
\pagestyle{plain} %

\title{
ADEPT-Z: Zero-Shot Automated Circuit Topology Search for 
\\Pareto-Optimal Photonic Tensor Cores
}

\author
{
Ziyang Jiang$^{1}$,
Pingchuan Ma$^{1}$,
Meng Zhang$^{2}$, 
Rena Huang$^{2}$,
Jiaqi Gu$^{1\dagger}$\\
$^{1}$Arizona State University, 
$^{2}$Rensselaer Polytechnic Institute\\
\small\textit{$^\dagger$jiaqigu@asu.edu}
}

\thispagestyle{empty}

\begin{abstract}
\label{abstract}
Photonic tensor cores (PTCs) are essential building blocks for optical artificial intelligence (AI) accelerators based on programmable photonic integrated circuits.
Most PTC designs today are manually constructed, with low design efficiency and unsatisfying solution quality. 
This makes it challenging to meet various hardware specifications and keep up with rapidly evolving AI applications.
Prior work has explored gradient-based methods to learn a good PTC structure differentiably. 
However, it suffers from slow training speed and optimization difficulty when handling multiple non-differentiable objectives and constraints.
Therefore, in this work, we propose a more flexible and efficient zero-shot multi-objective evolutionary topology search framework \name that explores Pareto-optimal PTC designs with advanced devices in a larger search space.
Multiple objectives can be co-optimized while honoring complicated hardware constraints.
With only $<$3 hours of search, we can obtain tens of diverse Pareto-optimal solutions, 100$\times$ faster than the prior gradient-based method, outperforming prior manual designs with 2$\times$ higher accuracy weighted area-energy efficiency.
The code of \name is available at \href{https://github.com/ScopeX-ASU/ADEPT-Z}{link}.
\end{abstract}

\maketitle
\section{Introduction}
\label{sec:Introduction}
Photonic tensor cores (PTC) offer significant advantages in artificial intelligence (AI) acceleration in terms of speed and energy efficiency over traditional electronic processors. 
Various integrated photonic tensor core designs have been demonstrated for speed-of-light matrix multiplication~\cite{NP_NATURE2017_Shen, NP_PIEEE2020_Cheng, NP_NaturePhotonics2021_Shastri, NP_ACS2022_Feng, NP_Science2024_Xu, NP_SciRep2017_Tait, NP_Nature2021_Xu, NP_Nature2021_Feldmann, NP_NatureComm2022_Zhu,NP_SciRep2024_zelaya,NP_NatureComm2021_Zhang,NP_APLML2024_Gu}.
Based on the physical principle of computing, PTCs can be classified into coherent and incoherent designs.
Coherent PTCs leverage phases of the light to encode more information and perform linear transformation via light interference.
The transfer matrix of coherent PTCs is usually a complex-valued matrix with stronger expressivity than real-valued tensor cores~\cite{NP_NatureComm2021_Zhang,NP_APLML2024_Gu}.
Based on the matrix expressivity, coherent PTCs can be further separated into \textbf{universal} PTCs that can realize arbitrary complex matrices and \textbf{subspace} PTCs whose implementable matrices are a subset of all matrices.
Clements/Reck-style Mach-Zehnder interferometer (MZI) meshes belong to universal PTCs.
Using singular value decomposition (SVD), two unitary matrices and a diagonal matrix can constitute arbitrary linear transformation.
Extensive subspace coherent PTCs have been proposed to increase hardware efficiency and scalability.
Butterfly-style PTC~\cite{NP_ASPDAC2020_Gu, NP_TCAD2020_Gu, NP_ACS2022_Feng} has been proposed to reduce the high cost of unitary matrices by using logarithmic-depth butterfly photonic mesh.
Interlacing MZI mesh based on repeated phase shifters and multi-port couplers~\cite{NP_SciRep2024_zelaya} has been proposed for more robust programmable PTCs.

However, almost all PTCs available today are manually designed based on matrix decomposition, human intuition, or inspiration from signal processing, which only covers several points in the enormous design space.
Universal MZI arrays have maximum expressivity but suffer from a large area and huge insertion loss.
Butterfly mesh is very compact in size, but its expressivity is limited when scales to larger arrays.
Besides those two extreme points, most space has been left unexplored.
Designing Pareto-optimal PTCs that honor multiple constraints remains a significant challenge due to the complex trade-offs among performance metrics, especially when it scales to large circuit sizes.
Even an experienced researcher often requires huge design efforts to create a photonic circuit design that can simultaneously deliver high matrix expressivity, high machine learning (ML) task accuracy, low latency, small area, and low power.
It is promising to develop an automated circuit topology search methodology to explore the design space of PTCs to push the Pareto-front in the accuracy/area/efficiency space with fast design closure.
Prior work has formulated the PTC circuit topology search as a differentiable optimization problem and used a gradient-based method to perform a one-shot topology search.
Parameters and architecture variables are co-optimized on a certain model and dataset.
This method successfully finds PTC designs with higher accuracy and smaller device footprint than MZI arrays and butterfly mesh.
However, it shows several key limitations.
(1) \textbf{The gradient-based circuit search method is limited to differentiable objectives}.
It takes considerable effort to mathematically relax the combinatorial optimization problem to its continuous equivalence.
However, not all objectives, such as the longest path, bounding box, or sequence distance, can be converted to a differentiable version.
Finding an accurate approximation and effective proxy also takes non-trivial efforts.
Moreover, the differentiable formulation restricts the search space such that it \emph{cannot consider multi-port optical couplers or arbitrary coupler placements} in the circuits.
(2) \textbf{It is difficult to handle multiple constraints}.
Since many circuit constraints are non-differentiable in nature, they often need to be gradually enforced by using penalty or Lagrangian methods.
Too many penalty terms make it difficult to balance their gradients and, thus, hard to converge to a high-quality and feasible solution.
(3) \textbf{High search cost to explore the Pareto front}.
One-shot gradient-based PTC search uses a weighted sum to optimize a single objective, which converges to one solution point after hours of network training time.
The search process needs to be relaunched every time the constraints or objective preference (i.e., weighting coefficients for metrics) change.

Motivated by the above limitations, in this work, we propose an efficient and flexible zero-shot PTC topology search framework \name based on gradient-free multi-objective evolutionary search, co-exploring the Pareto frontier with multiple objectives and hardware constraints in a larger design space.
Our main contributions are as follows:
\squishlist
    {\item We introduce a zero-shot topology search framework to explore Pareto-optimal photonic tensor core designs automatically.}
    {\item \textbf{Larger Design Space}: We expand the design space to include advanced multi-port devices with arbitrary placements for more efficient information interaction.}
    {\item \textbf{Multi-Objective Optimization}: We create a compact gene encoding with customized mutation/crossover operators for evolutionary search with balanced exploration and exploitation, which generates diverse Pareto-optimal solutions in accuracy-density-efficiency space, honoring area, power, and latency constraints.}
    {\item \textbf{Efficient Zero-Shot Performance Evaluation}:
    To avoid the high training cost, we introduce a comprehensive accuracy proxy based on efficient circuit trainability and expressivity evaluation.
    Rigorous layout area and circuit power calculations are used to evaluate efficiency and compute density. 
    }
    {\item Extensive evaluation on various benchmarks and circuit scales demonstrate that our searched PTC topologies show superior accuracy, compute density, energy efficiency, and generalizability with >100$\times$ lower search runtime compared to manual designs and prior auto-design method.}
\squishend

\section{Background}
\label{sec:Background}
\subsection{Automated PTC Design}
Previously, a differentiable PTC design method \texttt{ADEPT}~\cite{NP_DAC2022_Gu} has been proposed to formulate the combinatorial circuit topology search as a continuous probabilistic optimization problem and solve it with gradient descent.
Discrete device placement and routing problems are re-formulated as binarization-aware training and permutation matrix learning.
Expressivity of the PTC is optimized by training the constructed optical neural network (ONN) on a small dataset.
During training, hardware constraints and device footprint constraints are gradually enforced by the penalty method and augmented Lagrangian method.
This method takes 6-10 hours time to converge to a single feasible solution with carefully balanced objectives and penalty terms, which lacks flexibility for multiple objectives and complicated constraints and search efficiency to explore the Pareto frontier. 

\section{Zero-Shot Automatic Photonic Tensor Core Design Framework}
\label{sec:Method}
\subsection{Search Space Specification}
\label{sec:SearchSpaceSpec}
\begin{figure}
    \centering
    \includegraphics[width=1\columnwidth]{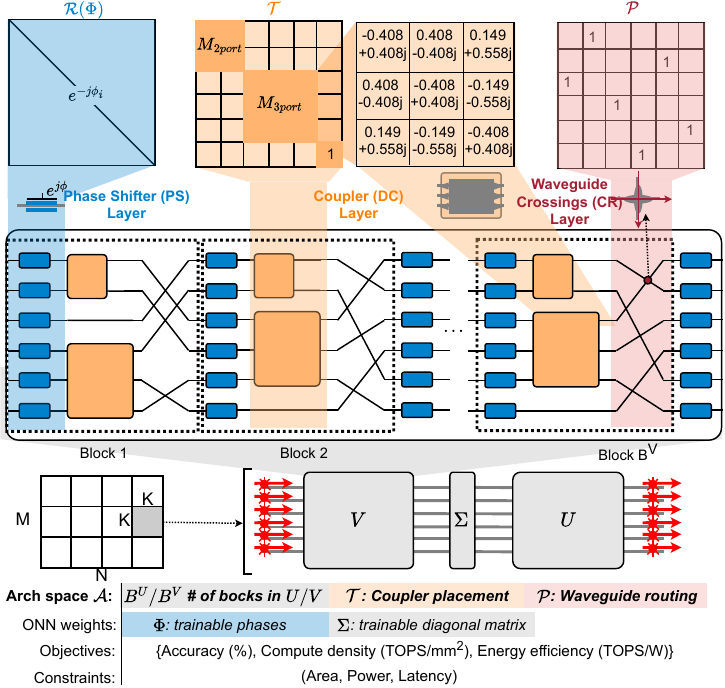}
    \vspace{-10pt}
    \caption{Illustration of PTC search space of \name.}
    \label{fig:Design_Space}
    \vspace{-10pt}
\end{figure}

For a complex-valued weight matrix $W\in\mathbb{C}^{M\times N}$, we can partition it into $P\times Q$ sub-matrices with the size of $K\times K$.
Each submatrix block can be mapped to a size-$K$ PTC.
Our goal is to search the Pareto-optimal topology for this $K \times K$ PTC.
As shown in Fig.~\ref{fig:Design_Space}, we adopt $U\Sigma V$ as the design skeleton.
Both of the unitaries $U_{pq}^{\alpha}$ and $V_{pq}^{\alpha}$ follow a pre-defined block-wise structure, each block containing a column of phase shifters $\mathcal{R}$, couplers $\calT$, and waveguide crossings $\calP$.
The diagonal matrix $\Sigma$ is simply a column of modulators.
Their transfer matrices can be formulated as 
\begin{equation}
\small
    U_{p q}^\alpha=\prod_{b=1}^{B^U} \mathcal{P}_b \mathcal{T}_b \mathcal{R}\left(\Phi_{p q}^b\right), \quad V_{p q}^\alpha=\prod_{b=B^U+1}^{B^U+B^V} \mathcal{P}_b \mathcal{T}_b \mathcal{R}\left(\Phi_{p q}^b\right)
\end{equation}
For simplicity, we only discuss $U$ and use $B$ instead of $B^U$.

\textbf{The first stage} of each block is one column of $K$ phase shifters (PS), which is equivalent to a diagonal matrix $\mathcal{R}\left(\Phi_{p q}^b\right)$ to the input vectors
$\mathcal{R}\left(\Phi_{p q}^b\right)=\operatorname{diag}\left(e^{-j \phi_1}, \cdots, e^{-j \phi_K}\right)$.
\textbf{The second stage} consists of multi-port couplers (DC) for all-to-all information mixing via diffraction and interference. 
Specifically, the multi-port couplers are Multi-Mode Interference (MMI) couplers. 
The transmission from $k$-th input port to $l$-th output port of an $N_c$-port general MMI~\cite{NP_Jlt2012_Zhou} is
\begin{equation}
\small
\begin{aligned}
& M_{l k}=(-1)^{l+k} j \exp \left(j \frac{\pi}{4}\right) \\
& \quad \times \sqrt{\frac{1}{N_c}} \exp \left(-j\left((l-1 / 2)-(-1)^{l+k}(k-1 / 2)\right)^2 \pi/(4N_c)\right),
\end{aligned}
\end{equation}
where $N_c$ is the number of the input/output ports.
An array of DCs (or waveguides) is expressed as a block diagonal matrix $\mathcal{T}_b$.
Our design space for each DC layer is equivalent to partitioning an integer $K$ into the sum of $n\in[1,K]$ nonnegative integers times their permutations, which is considerably larger than only densely placing 2-port DCs to fill all $K$ wires in prior work~\cite{NP_DAC2022_Gu}.
Since there is no analytical form for the combinations, we denote the exponential solution space for each DC layer as $F_K$.
\textbf{The last stage} in the block is the waveguide crossings (CR) that perform arbitrary bijective waveguide routing. 
The waveguide crossing layer can be expressed as a permutation matrix $\mathcal{P}_b$. 
The $\mathcal{P}_b$, which has $K!$ possible combinations in one block, contributes most of the design space.

\noindent\textbf{\underline{Design Space}}.~
In summary, a photonic mesh contains \(B\) blocks, each comprising a PS layer, a DC layer, and a CR layer. 
The topology \(\alpha\) includes the number of blocks \(B^U\) and \(B^V\), the waveguide connections $\calP$, and the placements of couplers as specified by $\calT$. 
The total design space is $O\left((F_K \cdot K!)^{B_{\max }}\right)$. 

\subsection{Problem Formulation}
Our goal is to explore the Pareto-front of coherent PTC designs to deliver high expressivity, area efficiency, and energy efficiency while honoring the area, power consumption, and latency constraints. 
We formulate the constrained multi-objective problem as follows, 
\begin{equation}
\small
\label{eq:Formulation}
\begin{gathered}
\max _{\alpha \in \mathcal{A}} \{\mathcal{S}_{1}\left(gW^{*\alpha} \right), CD\left(\alpha \right),
EE\left(\alpha \right) \},  \quad 
\alpha=\left(B^U, B^V, \mathcal{P}, \mathcal{T}\right) \\
\text { s.t. } W^*=\underset{W}{\operatorname{argmin}} \mathcal{L}\left(W^\alpha ; \mathcal{D}^{\text {trn }}\right), C_{\text {min }_i} \leq \mathcal{C}_i(\alpha) \leq C_{\text {max }_i}, i \in \mathbb{N^+} \\
W^\alpha \in \mathbb{C}^{M \times N}=\left\{W_{p q}^\alpha\right\}_{p=1, q=1}^{p=P, q=Q}=\left\{U_{p q}^\alpha \Sigma_{p q} V_{p q}^\alpha\right\}_{p=1, q=1}^{p=P, q=Q}, \\
B^U, B^V \in\left[B_{\text {min }} / 2, B_{\text {max }} / 2\right], W_{p q} \in \mathbb{C}^{K \times K}, \\
\mathcal{P}=\left(\cdots, \mathcal{P}_b, \cdots, \mathcal{P}_{B^U+B^V}\right), \mathcal{T}=\left(\cdots, \mathcal{T}_b, \cdots, \mathcal{T}_{B^U+B^V}\right),
\end{gathered}
\end{equation}
where $W^*\in\mathbb{C}^{M\times N}$ is the trained ONN weight matrix for accuracy evaluation.
There are multiple hardware constraints $C_i$ that need to be honored.
The main optimization variables are architecture parameters $\alpha$ that impact the structure of $U$ and $V$ circuits, which contains circuit block count $B^{U/V}$, coupler layer transmission $\calT$, and waveguide crossings $\calP$.
Diagonal $\Sigma$ belongs to ONN weights, not circuit topology.
The PTC topology \textbf{$\alpha$ is shared} for all matrices in the neural network layers, not layer-specific.

\subsection{Multi-Objective Evolutionary PTC Search}
\begin{algorithm}
    \small
    \caption{Evolutionary PTC topology search algorithm}
    \label{alg:genetic}
    \begin{algorithmic}[1]
        \Require{Maximum iteration times $I_{max}$, Second phase iteration time $I_{phase2}$, Population size $P_0$, Initial mutation rate $p_{mu}$, constant crossover rate $p_{co}$, Baseline topologies $\alpha_0$}
        \Ensure{Optimal solution set $S$}
        \State $S \gets \alpha_0 + \texttt{randomInit}(P_0)$ \Comment{Section 3.3.2, 3.3.3}
        \For {$i \gets 1\cdots I_{max}$}
            \State OffSpring $ \gets \texttt{mutateAndCrossover}(S, p_{mu}, p_{co})$\Comment{Section 3.3.4}
            \State OffSpring $\gets \texttt{checkConstraints}(\text{OffSpring})$\Comment{Section 3.3.5}
            \State $S\gets S + \text{OffSpring}$
            \State $S.\texttt{assessSortSelect}()$\Comment{Section 3.3.1, 3.3.6}
            \If {$i \leq I_{max} - I_{phase2}$}\Comment{Section 3.3.6}
                \State $p_{mu} \gets \texttt{cosineDecayScheduler.step}()$
            \EndIf
        \EndFor
    \end{algorithmic}
\end{algorithm}

To explore the \emph{highly discrete} and \emph{multi-objective} PTC topology design space, a suitable search engine needs to meet the following requirements: 
\ding{202} it can handle multiple (non-)differentiable objectives; 
\ding{203} it should ensure sufficient exploration with balanced exploitation to cover the huge design space and generate multiple Pareto-optimal candidate solutions; 
\ding{204} it must ensure feasibility under multiple constraints;
\ding{205} it should be efficient, especially avoiding high model training cost to find the trained weights $W^*$.
To satisfy all the above requirements, we propose a zero-shot gradient-free framework based on a customized multi-objective evolutionary search algorithm to efficiently solve the \textbf{constrained combinatorial optimization problem with multiple objectives} as shown in Alg.~\ref{alg:genetic}.

\vspace{-5pt}
\subsubsection{Objective Definition}
\label{sec:ObjectiveDefinition}
Multiple metrics must be considered in a balanced fashion when assessing a PTC topology. 
\textit{We employ three scores simultaneously to evaluate the performance, including task accuracy, compute density, and energy efficiency.}

\begin{figure}
    \centering
    \vspace{-5pt}
    \includegraphics[width=1\columnwidth]{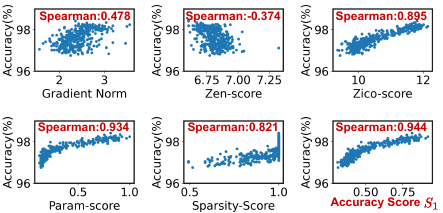}
    \vspace{-10pt}
    \caption{Spearman coefficients for different accuracy proxies.}
    \label{fig:Acc_Score_Pick}
    \vspace{-10pt}
\end{figure}

\noindent\textbf{\underline{Accuracy Score}}.~
The accuracy impact of a PTC needs to be measured by mapping it onto an ONN and training and evaluating the model on a dataset. 
To avoid training costs and expedite the search process, we design a comprehensive Accuracy Score as a training-free proxy, which evaluates both the trainability and expressivity of PTC topology.

We select 5 candidate scores to construct the proxy: (1) Param-Score, (2) Sparsity-Score, (3) Zico-Score~\cite{NAS_ICLR2023_Li}, (4) Zen-Score~\cite{NN_ICCV2021_Lin}, and (5) Gradient Norm~\cite{NAS_ICLR2021_abdelfattah}. 
The first two scores are motivated by the insight that a matrix's \textbf{expressivity} is typically related to the number of independent parameters and its sparsity.
For Param-Score, phases $\Phi$ on programmable PS are the only trained parameters in unitaries.
We count the total PS (directly connected phase shifters can be merged as one) normalized to the matrix size $K^2$ as the Param-Score.
Even with many parameters, if there is insufficient cross-channel interaction via couplers, the matrix can be a block diagonal matrix with high sparsity.
Hence, as a complementary score, we use the sparsity of $W$ (higher is denser) to evaluate its expressivity.

The last three are commonly used accuracy proxies from Zero-shot neural architecture search (NAS), which all focus on the gradient/Lipschitz-related property to evaluate its \textbf{trainability}.
Figure~\ref{fig:Acc_Score_Pick} shows the Spearman correlation between each score and the test accuracy obtained from extensive training to measure the ability of the scores to accurately predict the \emph{relative accuracy ranking} of different PTC topologies. 
\emph{Zico-Score, Param-Score, and Sparsity-Score} have the highest Spearman coefficients, indicating more accurate predictions. 
We define an Accuracy Score $\mathcal{S}$ as a linear weighted combination of these three scores. 
Optimal combination coefficients are found by solving the following optimization problem to maximize the Spearman correlation,
\begin{equation}
\small
\begin{aligned}
    \max _{c_i} \texttt{Spearman}(\mathcal{S}(\alpha), Acc(W^{*\alpha})), \mathcal{S}(\alpha) = \sum c_i*S_i, \\
    S_i \in \{\text{Zico-Score, Param-Score, Sparsity-Score}\}
\end{aligned}
\end{equation}
in which $Acc(W^{*\alpha})$ is the actual test accuracy of a trained ONN model. 
The final Accuracy Score $\calS({\alpha})$ is given: 
\begin{equation}
\begin{gathered}
\calS({\alpha}) = 0.015 \cdot S_{\text{Zico}}({\alpha}) + 0.561 \cdot S_{\text{Param}}({\alpha}) + 0.175 \cdot S_{\text{Sparsity}}({\alpha})
\end{gathered}
\end{equation}

\noindent\textbf{\underline{{Compute Density (CD)}}}.~
Compute density (CD) is a commonly used performance metric in AI accelerators, which measures the computing speed with a unit chip area, typically in the unit of TOPS/mm$^2$.
Higher CD means better area efficiency. 
For a given topology $\alpha$, the compute density $CD({\alpha})$ is as follows:
\begin{equation}
\begin{gathered}
\text{CD}({\alpha}) = 2K^2/(A({\alpha}) \times \tau({\alpha})) 
\end{gathered}
\end{equation}
A({$\alpha$}) represents the estimated area of $\alpha$ and $\tau$({$\alpha$}) represents the estimated latency of the PTC. 
Later, we will give a detailed estimate of the PTC area and latency.

\noindent\textbf{\underline{Energy Efficiency (EE)}}.~
Energy efficiency (EE), typically in the unit of TOPS/Watt, is a vital objective for efficient AI hardware. 
The formula to calculate energy efficiency $EE({\alpha})$ is as follows:
\begin{equation}
\begin{gathered}
\text{EE}({\alpha}) = 2K^2/(P({\alpha}) \times \tau({\alpha})), 
\end{gathered}
\end{equation}
where $P(\alpha)$ represents the estimated power explained later.

For each potential solution, we evaluate these three objectives independently. The solutions are then ranked and selected based on their combined performance across Accuracy Score, Compute Density, and Energy Efficiency. 
Instead of a weighted sum of those three objectives with heuristic preference, we \textbf{simultaneously maximize three scores} to obtain multiple Pareto-optimal points, from which designers can further select suitable designs by only searching once.

\subsubsection{Gene Encoding}
\label{sec:GeneEncoding}
\begin{figure}
    \centering
    \vspace{-5pt}
    \includegraphics[width=0.95\columnwidth]{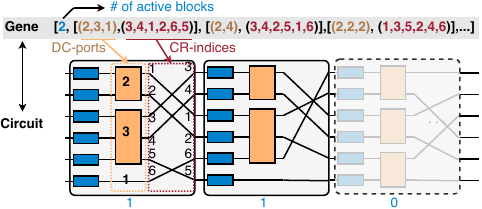}
    \vspace{-10pt}
    \caption{Gene-to-circuit mapping.}
    \label{fig:Encoding}
    \vspace{-5pt}
\end{figure}

To facilitate evolutionary search, we create a compact gene representation to encode the topology of a PTC in Fig.~\ref{fig:Encoding}.
The gene starts with a number $B$ to indicate the first $B$ blocks are active, followed by $B^{max}$ block encodings, each carrying multi-port DC placement information and waveguide permutation indices.

\noindent\textbf{\underline{Block Encoding}}.~
The first integer in a gene is the number of effective blocks in $\alpha$.
The first $B/2$ blocks construct $U$, while the rest are for $V$.

\noindent\textbf{\underline{DC Array Encoding}}.~
For the DC array, we use a sequence of nonnegative integers $\{(N_c^1,\cdots, N_c^{n})|\sum_{i=1}^nN_c^i=K, N_c\in\mathbb{N}_+\}$, where each integer corresponds to a specific type of multi-port DC ($N_c>1$) or a waveguide ($N_c=1$). 
For instance, a 3-port DC is denoted as 3. 
This approach simplifies the representation of the DC array, making it easy to parse and manipulate during the evolutionary process.

\noindent\textbf{\underline{CR Array Encoding}}.~
The permutation indices, i.e., positions of '1' in the permutation matrix $\calP$, are compact representations or waveguide routing solutions.
All feasible solutions can be efficiently accessed by re-ordering the indices.

\vspace{-5pt}
\subsubsection{Population Initialization}
\label{sec:Initialization}
The population size is $P_0$, and we randomly sample the number of active blocks $B$ and DC placements for each population.
Randomly permuted indices for CR layers have too many crossings. 
Thus, we heuristically limit the maximum crossings for each CR layer not to exceed the maximum crossings in butterfly mesh, i.e., $K(K/2-1)/4$.
Initialized populations will honor all hardware constraints.
Manual designs have also been added as initial solutions.

\vspace{-5pt}
\subsubsection{Mutation \& Crossover}
\label{sec:MutationCrossover}
We customize global/local mutation and crossover operators to ensure better global coverage of the large design space while facilitating better convergence with local search.

\noindent\textbf{\underline{Mutation Operator}}.~
We designed three types of mutation operators based on the type of devices they apply to, summarized in Table~\ref{tab:MutationOperators}. 
For \textbf{DC Mutation}, we design 4 operators: (A2R1, R2A1, Move) for local adjustment, and RS to escape the local optima. 
For \textbf{CR Mutation}, we design 2 operators: AddCR and ReduceCR. 
Applying $\Delta N_c$ steps of bubble sort (descending) to the CR-indices is equivalent to reducing the same amount of crossings.
Ascending sort has the opposite effect of increasing crossings.
The maximum crossings are still limited by the value in butterfly meshes, as explained during population initialization. 
For \textbf{Block Mutation}, we have 2 operators: AddBlock and ReduceBlock, which mainly adjust the circuit depth. 
For DC and CR arrays, before applying mutation, we first perform a \textbf{legality check} to ensure that the operator can be successfully applied to the gene segment. 
Then, with a mutation probability $p_{mu}$, we randomly select one legal operator from the operator set and apply it to the gene.

\begin{figure}
\vspace{-10pt}
    \centering
    \subfloat[]{\includegraphics[width=0.38\columnwidth]{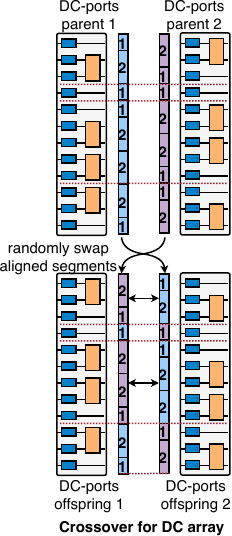}
    \label{fig:DCCrossover}}
    \hspace{5pt}
    \subfloat[]{\includegraphics[width=0.56\columnwidth]{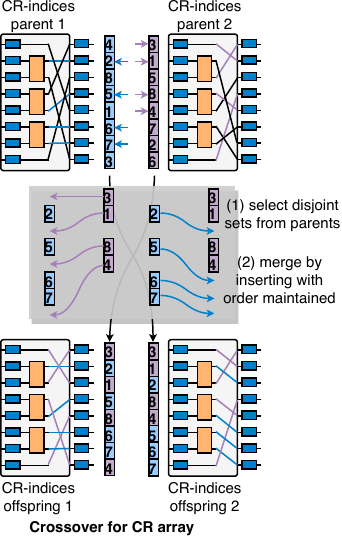}
    \label{fig:CRCrossover}}
    \vspace{-10pt}
    \caption{Crossover for (a) DC and (b) CR arrays.}
    \label{fig:Crossover}
    \vspace{-10pt}
\end{figure}

\noindent\textbf{\underline{Crossover Operator}}.~
We customize crossover operators for solution interpolation, shown in Fig.~\ref{fig:Crossover}. 
For \textbf{DC Crossover}, we identify all potential cutting points to divide parent genes into \emph{border-aligned} segments, which avoids cutting through multi-port couplers. 
Sliced segments are swapped with a probability of 0.5. 
For \textbf{CR Crossover}, to ensure \emph{legal} indices while \emph{preserving the relative order} in parent genes, we select even-sized disjoint indices from two parents, shown in Fig.~\ref{fig:CRCrossover}.
Then, we insert the selected indices from the other parent into the empty slots of one parent and generate two offspring.
For \textbf{Block Crossover}, we randomly swap two active(effective) blocks at the same position with a probability of 0.5 to avoid generating illegal genes. 
All the detailed crossover methods are illustrated in Fig.\ref{fig:Crossover}.

\begin{table}[]
\centering
\caption{Mutation operators for different components.}
\vspace{-10pt}
\label{tab:MutationOperators}
\large
\resizebox{1.05\columnwidth}{!}{%
\begin{tabular}{c|c|p{10cm}}
\toprule
\textbf{Type} & \textbf{Mutation Ops} & \textbf{Description} \\ \midrule
\multirow{4}{*}{DC} 
& R2A1 & Remove 2 and add 1 DC. E.g., [2,4,2] $\rightarrow$ [1,1,1,1,1,1,2] $\rightarrow$ [1,4,1,2] \\ \cmidrule{2-3} 
& A2R1 & Add 2 and remove 1 DC. E.g., [1,1,2,1,1,1,1] $\rightarrow$ [1,1,2,2,2] $\rightarrow$ [1,1,2,1,1,2]  \\ \cmidrule{2-3} 
& Move & Move one DC to another position. E.g., [1,2,2,2,1] $\rightarrow$ [1,2,2,1,2] \\ \cmidrule{2-3} 
& RS   & Resample a DC array \\ \midrule

\multirow{2}{*}{CR} 
& AddCR & Add a random number of CR. E.g., Add 2 CRs: [0,1,2,3] $\rightarrow$ [0,3,1,2] \\ \cmidrule{2-3} 
& ReduceCR & Reduce a random number of CR. E.g., Reduce 1 CR: [0,2,1,3] $\rightarrow$ [0,1,2,3]\\ \midrule

\multirow{2}{*}{Block} 
& AddBlock & Copy a random number of blocks from the front of gene, add to the end \\ \cmidrule{2-3} 
& ReduceBlock & Remove a random number of blocks from the end of gene. \\ \bottomrule
\end{tabular}
}
\end{table}

\vspace{-5pt}
\subsubsection{Cost Estimation}
\label{sec:EvaluationOfAreaPowerLatency}
Here, we explain a detailed estimation of hardware cost, including area, power, and latency.

\begin{figure}
    \centering
    \includegraphics[width=0.8\columnwidth]{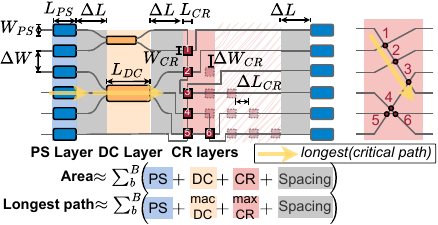}
    \vspace{-10pt}
    \caption{Layout for area/latency estimation. 
    A compact crossing array layout that occupies the leftmost slots is pre-defined.}
    \label{fig:Cost_Estimation}
    \vspace{-5pt}
\end{figure}

\noindent\textbf{\underline{Area Estimation}}.~
For a $K\times K$ PTC, we estimate the area cost of all its electrical and optical components as follows:
\begin{equation}
\small
\label{eq:Area}
\begin{aligned}
A({\alpha}) \!\!=& A_{U}({\alpha})+A_{V}({\alpha}) \!+\! A_{\Sigma}\!+\! K(A_{\text{TIA}}\!+\! A_{\text{PD}}\!+\!A_{\text{MZM}}\!+\! A_{\text{DAC}}\!+\!A_{\text{ADC}}),
\end{aligned}
\end{equation}
where $A_{\text{TIA}}$, $A_{\text{PD}}$, $A_{\text{MZM}}$, $A_{\text{ADC}}$ and $A_{\text{DAC}}$ are area cost for trans-impedance amplifier (TIA), photodetector (PD), high-speed Mach-Zehnder modulator (MZM), analog-to-digital (ADC), and digital-to-analog converter (DAC). 
The area for the photonic part is:
\begin{equation}
\small
\label{eq:USVArea}
\begin{aligned}
A_{U/V}&({\alpha}) = L_{PS}(W_{PS} + (K-1) \Delta W)
+ L_{DC}(K-1) \Delta W \\
&+ \big(N_c L_{CR} + (N_c-1) \Delta L_{CR}\big)\big(N_r W_{CR} + (N_r-1) \Delta W_{CR}\big) \\
&+ \big(3(K-1)\Delta W\Delta L + W_{PS}\Delta L\big),\\
A_{\Sigma}& = ((2K-1)\Delta W + W_{PS})\cdot(L_{PS}+2\Delta L)\\
&+((K-1)\Delta W+L_{Y})(2L_{Y}+\Delta W),
\end{aligned}
\end{equation}
where $L$ and $W$ represent device length and width for phase shifter (PS), coupler (DC), crossing (CR), and y-branch (Y).
$\Delta L$ and $\Delta W$ are spacings. 
$N_c$ and $N_r$ represent the number of columns and rows occupied by our predefined compact triangular crossing array layout, which fills the leftmost column first and expands to the right.
Figure~\ref{fig:Cost_Estimation} shows the details for estimating the hardware cost of the unitary matrices. 
\textbf{Our area estimation considers the actual chip layout and practical spacing, which is much more accurate than simply summing up all device footprint in prior work~\cite{NP_DAC2022_Gu}}.
All dimensions for optical components can also be obtained from the GF foundry PDK. 
We set $\Delta L=20\mu m$, $\Delta W=100\mu m$, $W_{CR}=L_{CR}=10\mu m$.

\noindent\textbf{\underline{Power Estimation}}.~
\label{sec:PowerEstimation}
The PTC power is estimated as follows~\cite{NP_ACS2022_Feng}:
\begin{equation}
\small
\begin{gathered}
P({\alpha}) = P_{\text{laser}} + K \cdot (P_{\text{MZM}} + P_{\text{DAC}} + P_{\text{ADC}} + P_{TIA} +  P_{\text{PD}}).
\end{gathered}
\end{equation}
The formula for laser power is: $P_{\text{laser}}= \frac{2^{b}\cdot 10^{(S_{PD} + IL)/10}}{\eta}$, where $\eta$ is the wall-plug efficiency, 
$S_{PD}$ is the PD sensitivity,
IL is the insertion loss of the circuit, and $b$ refers to the ADC bit resolution. 
Given working frequency $f$ and input bitwidth $b$, the power for DAC is derived by: $P_{DAC}= \frac{b_{0}2^{b}f}{b2^{b_{0}}f_{s}} \cdot P_{DAC0}$, where $P_{DAC0}$ is the power of DAC at sampling rate $f_{s}$ and $b_{0}$ bit precision. 
The ADC power is derived by: $P_{ADC}= \frac{b_{0}f}{bf_{s}} \cdot P_{ADC0}$, where $P_{ADC0}$ is ADC power at sampling rate $f_{s}$ and $b_{0}$ bit precision. 
$P_{\Sigma}$ and $P_{PS}$ consider the static power of all phase shifters.

\noindent\textbf{\underline{Latency Estimation}}.~
\label{sec:LatencyEstimation}
The PTC latency is determined by the optical path delay, input modulation, and readout delay as follows: 
\vspace{-3pt}
\begin{equation}
\small
\begin{aligned}
\tau({\alpha}) = \max(f^{-1}, n_{\text{g}} L_{\text{path}}/c_{0} + \tau_{DAC} + \tau_{PD}),\\
L_{\text{path}}=\sum_{b}^B(L^b_{PS}+L^{b,max}_{DC}+L^{b,max}_{CR}+3\Delta L),
\end{aligned}
\vspace{-5pt}
\end{equation}
where the clock rate is set to $f$=10 GHz, $c_0$ is the light speed, $n_{\text{g}}$ is the group index, $L_{\text{path}}$ refers to the longest optical path length, and $\tau_{DAC}$ and $\tau_{PD}$ refer to the delays of DAC and PD, and they are both set to 10ps. 
For the estimation of the longest path length $L_{\text{path}}$, we consider the worst-case scenario. 
We sum up the PS length, largest coupler length, longest waveguide routing path, and $\Sigma$ matrix length to get $L_{\text{path}}$. 
If the optical path delay of a very deep circuit cannot be hidden by one cycle ($f^{-1}$), the clock frequency will be reduced to accommodate the latency accordingly~\cite{NP_HPCA2024_Zhu}.

\subsubsection{Two-stage Pareto Front Search Strategy: NSGA-II}~
\label{sec:SelectionStrategy}
We adopt a Non-dominated Sorting Genetic Algorithm II (NSGA-II) algorithm~\cite{NN_TEC2002_Deb} as the search engine to handle multi-objective optimization.
Each search iteration doubles the population size by generating new legal solutions via crossover and mutation, and then it selects solutions from the superior Pareto fronts that contribute the most to solution diversity to maintain a constant population size. 

To \textbf{prioritize exploration at the beginning and gradually focus on local exploitation}, 
we divide the search process into two phases.
In the first phase, we use all mutations for DC, CR, and Blocks with a cosine-decayed mutation rate, which allows us to explore better genes across the search space while ensuring that high-quality genes do not undergo significant mutations.
In the second phase, we set all mutation rates to 0.02 and remove the two block-level mutation operators and the RS operator for DC to avoid major gene changes. 
This ensures that only minor mutations occur for local search toward optimal solutions.

\begin{figure}
    \centering
    \vspace{-5pt}
    \includegraphics[width=0.95\columnwidth]{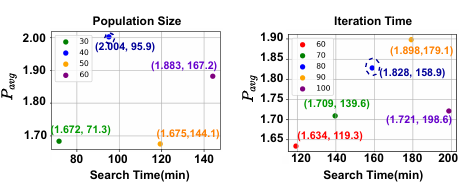}
    \vspace{-10pt}
    \caption{Population size of 40 and Iteration time of 80 achieve a balance between exploration and efficiency.
    }
    
    \label{fig:Exp1_Population}
    \vspace{-13pt}
\end{figure}

\begin{figure}
    \centering
    \includegraphics[width=1\columnwidth]{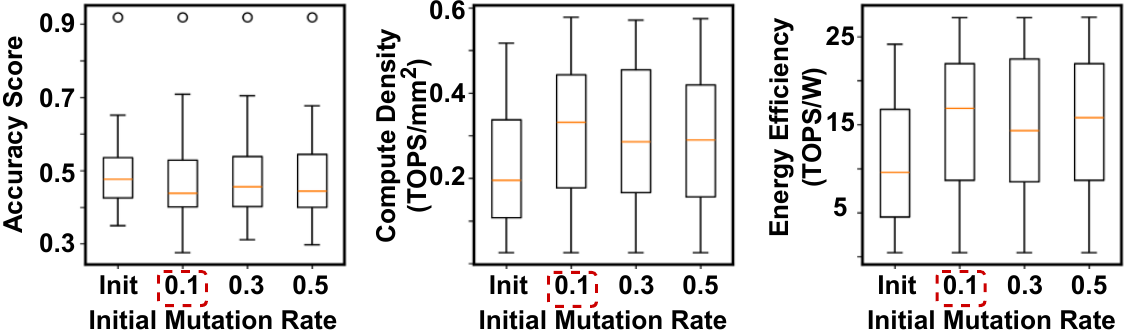}
    \vspace{-15pt}
    \caption{Initial mutation rate of 0.1 gives a good exploration.}
    \label{fig:Exp3_Initial_Value}
    \vspace{-10pt}
\end{figure}

\section{Experimental Results}
\label{sec:ExperimentalResults}
\subsection{Experiment Setup}
\label{sec:ExpSetup}
\begin{figure}
    \centering
    \vspace{-10pt}
    \subfloat[]{\includegraphics[width=1\columnwidth]{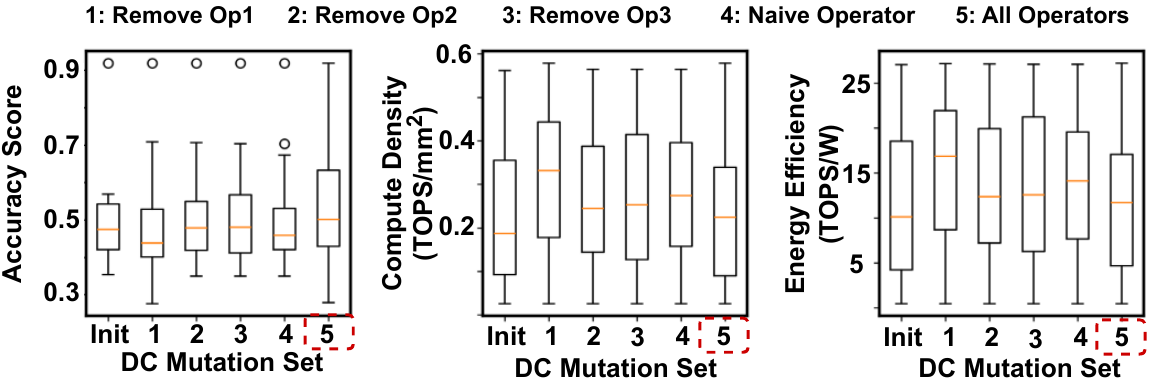}
    \label{fig:Exp4_Mutation_op_dc}}\\
    \vspace{-5pt}
    \subfloat[]{
     \includegraphics[width=1\columnwidth]{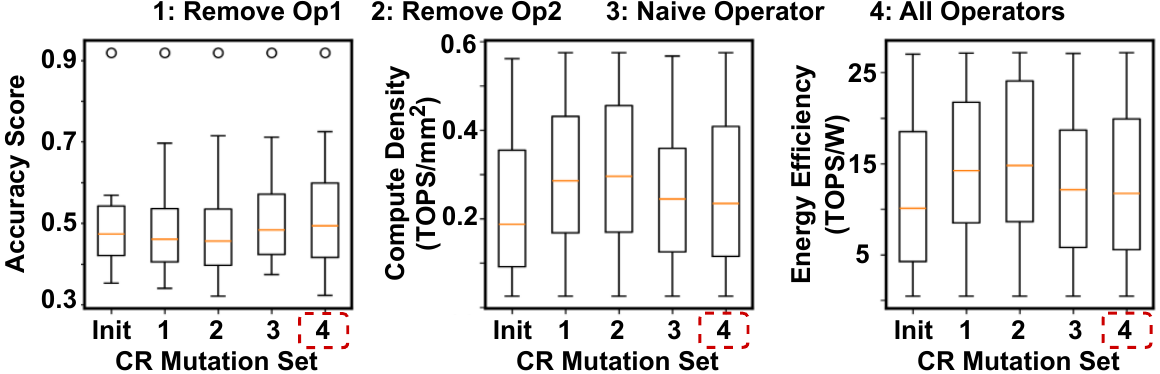}
     \label{fig:Exp4_Mutation_op_cr}}
    \vspace{-10pt}
    \caption{Validate each mutation operator for (a) DC and (b) CR array. We remove one mutation operator at a time and observe the distribution of the objective values of the final populations.}
    \vspace{-10pt}
\end{figure}

\noindent\textbf{\underline{Datasets}}.~During the search phase, we used the MNIST dataset to estimate Test Accuracy. 
The solutions found were then evaluated on the MNIST~\cite{NN_MNIST1998}, FMNIST~\cite{NN_FashionMNIST2017}, SVHN~\cite{NN_svhn2011}, and CIFAR10~\cite{NN_cifar2009} datasets.

\noindent\textbf{\underline{NN Models}}.~
The PTC topology is searched on a 2-layer CNN model and MNIST dataset without extensive training.
The searched topology is then applied to other models/datasets.

\noindent\textbf{\underline{Searching Settings}}.~
We choose the population size to be 40, the maximum iterations to be 80, and the initial mutation rate to be 0.1.
For 16$\times$16 PTCs, we use (2-port, 8-port) DCs during search. 
For $K$=8 and 32, we use (2-port, 4-port) for 8$\times$8 PTCs and (4-port,16-port) for 32$\times$32 PTCs. 
We applied area constraint [18.31,24.02] $mm^2$ (80\% of butterfly optical area up to 50\% of MZI array optical area, plus electrical area), power constraint [50, 1000] mW, latency constraint [100, 1000] ps to the search process. 
We set $f$=10 GHz and the resolution as 4-bit. 
For device cost, we use GF foundry PDK~\cite{NP_OFC2020_Rakowski} and a customized PDK~\cite{NP_JAP2024_Zhang}.
Electrical devices are the same as~\cite{NP_JAP2024_Zhang}.

\subsection{Ablation Studies}
\label{sec:PermutationAblation}
\noindent\textbf{\underline{Population Size and Search Steps}}.~
We use the product of three objectives to reflect the solution quality, and $P_{avg}$ is the average value across the current populations.
Figure~\ref{fig:Exp1_Population} shows that 40 populations evolved for 80 steps have the best quality and runtime balance.

\noindent\textbf{\underline{Mutation Rate and Operators}}.~
Figure~\ref{fig:Exp3_Initial_Value} determines the best initial mutation rate of 0.1 to balance exploration and convergence.
To verify the impact of the designed mutation operators, we removed one operator at a time, showing that all operators positively affect solution quality both for DC and CR, shown in Fig.~\ref{fig:Exp4_Mutation_op_dc} and Fig.~\ref{fig:Exp4_Mutation_op_cr}.

\begin{figure}
    \centering
    \includegraphics[width=0.99\columnwidth]{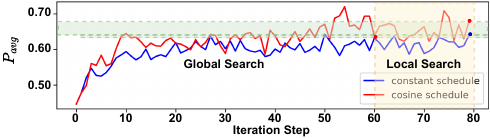}
    \vspace{-10pt}
    \caption{Compare the search performance of a constant scheduler and our two-stage cosine decay scheduler.
    }
    \label{fig:Exp9_Scheduler}
    \vspace{-5pt}
\end{figure}

\noindent\textbf{\underline{Two-Stage Search}}.~
Our two-stage search using the cosine mutation rate scheduler, as shown in Fig.~\ref{fig:Exp9_Scheduler}, converges faster to higher-quality solutions than the single-stage search with a constant mutation rate.

\subsection{Main Results}
\label{sec:MainResults}
\begin{figure}
    \centering
    \includegraphics[width=0.88\columnwidth]{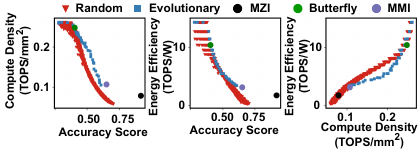}
    \vspace{-10pt}
    \caption{Comparison of the observed solution points between random search and evolutionary search. 
    }
    \label{fig:Pareto_Front}
    \vspace{-8pt}
\end{figure}

In Fig.~\ref{fig:Pareto_Front}, our solutions are Pareto-optimal as they dominate random designs and prior manual designs in the accuracy-density-efficiency space.
To better evaluate the performance, we introduce two comprehensive metrics: \textbf{Area-Energy Efficiency (AEE) and Accuracy-weighted AEE (AAEE), i.e., Accuracy-AEE product}.
From the final Pareto front, we select four designs for each PTC size, named \name-a0 to \name-a3, and compare them to manual designs, i.e., MZI array~\cite{NP_NATURE2017_Shen}, Butterfly mesh~\cite{NP_ACS2022_Feng}, and interlaced MMI array~\cite{NP_SciRep2024_zelaya} in Table~\ref{tab:CompareGF}. %

\name-a0 is the best solution in terms of AAEE on all three PTC sizes. 
Our solutions balance expressivity and hardware cost compared to MZI and MMI arrays, showing an overall 2.47$\times$ higher AAEE.
We observe that butterfly solutions are roughly located at the Pareto front.
Our solutions are 1.03$\times$ better than Butterfly in AAEE, but our method gives much more diverse designs to cover various accuracy/power/area/latency requirements.

\begin{figure}
    \centering
    \vspace{-5pt}
    \includegraphics[width=1\columnwidth]{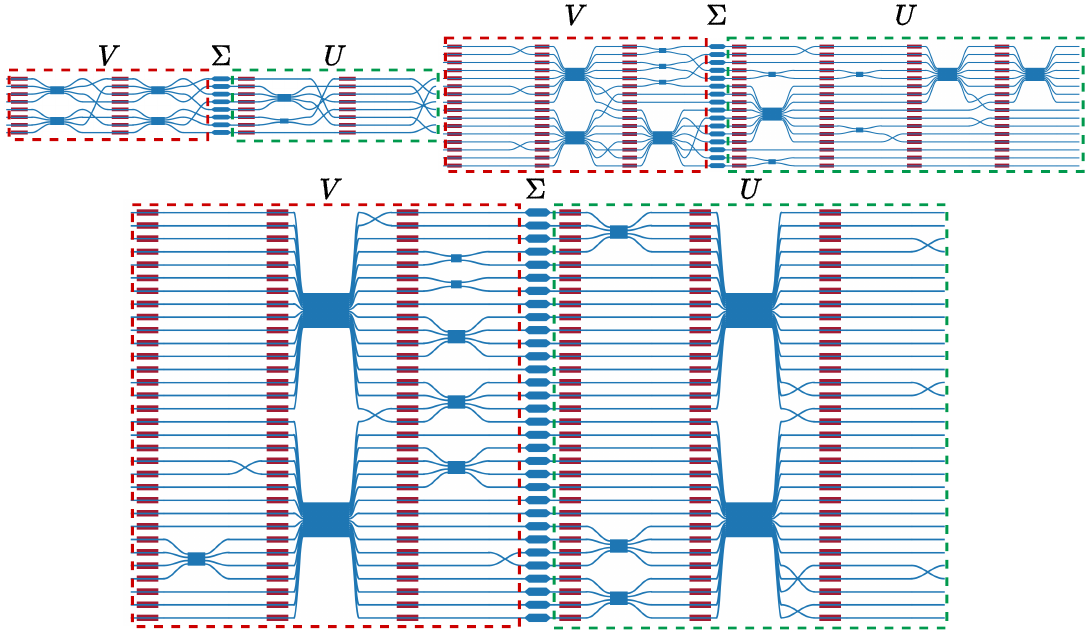}
    \vspace{-15pt}
    \caption{Visualization of \name-a0 for $K$=8, 16, and 32.
    }
    \label{fig:Visualization}
    \vspace{-5pt}
\end{figure}

We visualize \name-a1 in Fig.~\ref{fig:Visualization}. 
Multi-port DCs are frequently used for efficient cross-channel interaction.
As a result, waveguide crossings are minimized to mix signals only when necessary to reduce hardware costs.
Note that we do not compare to gradient-based \texttt{ADEPT}~\cite{NP_DAC2022_Gu} as \textbf{\texttt{ADEPT} cannot handle multi-port couplers or non-differentiable latency/area objectives}.
Moreover, our method can find 40 Pareto-optimal solutions within 2.7 hours, 100$\times$ faster than \texttt{ADEPT}, which requires 40$\times$8=320 hours even if it is applicable. 

\begin{table}[]
\centering
\caption{Evaluate PTCs with different sizes using GF PDK in terms of area (optical+electrical) (mm$^2$), power (mW), and latency (ps).
We also show compute density (CD) (TOPS/mm$^2$), energy efficiency (EE) (TOPS/W), area-energy efficiency (AEE) (TOPS/W/mm$^2$), and accuracy-weighted AEE (AAEE).
}
\vspace{-10pt}
\label{tab:CompareGF}
\resizebox{1.05\columnwidth}{!}{%
\begin{tabular}{cc|ccc|cccc}
\hline
$K$ & Metrics & MZI~\cite{NP_NATURE2017_Shen} & Butterly~\cite{NP_ACS2022_Feng} & MMI~\cite{NP_SciRep2024_zelaya} & \name-a0 & \name-a1 & \name-a2 & \name-a3 \\ 
                       \hline
\multirow{4}{*}{8}   & Area(O+E) & 3.79+8.18  & 0.92+8.18  & 3.57+8.18  & 0.73+8.18  & 0.83+8.18  & 0.94+8.18  & 1.57+8.18  \\
                              & Power  & $141.09$  & $141.48$  & $141.45$  & $141.92$  & $141.32$  & $141.98$  & $142.74$  \\                   
                              & Latency  & $100.69$  & $100.00$  & $100.00$  & $100.00$  & $100.00$  & $100.00$  & $100.00$ \\
                              & \text{Accuracy} & $98.68$  & $98.33$  & $98.72$  & $98.22$  & $98.13$  & $98.30$  &$98.38$  \\
                              
                       \hline
\multirow{4}{*}{16} & Area(O+E) & 15.32+16.36  & 2.43+16.36  & 21.16+16.36  & 2.37+16.36  & 2.39+16.36  & 3.38+16.36  & 3.44+16.36  \\
                              & Power  & $209.83$  & $283.15$  & $266.64$  & $282.19$  & $282.20$  & $284.18$  & $284.48$\\ 
                              & Latency  & $147.25$  & $100.00$  & $107.38$  & $100.00$  & $100.00$  & $100.00$  & $100.00$ \\
                              & \text{Accuracy} & $98.74$  & $98.16$  & $98.58$  & $97.83$  & $97.69$  & $98.27$  & $98.24$  \\
                              
                       \hline
\multirow{4}{*}{32} & Area(O+E) & 61.56+32.72  & 6.11+32.72  & 142.58+32.72  & 5.62+32.72  & 6.46+32.72  & 8.53+32.72  & 9.77+32.72  \\
                              & Power & $316.34$  & $487.36$  & $394.52$  & $563.50$  & $563.52$  & $563.69$  & $563.71$  \\            
                              & Latency& $240.37$  & $122.13$  & $158.26$  & $100.00$  & $100.00$  & $100.00$  & $100.00$  \\
                              & \text{Accuracy} & $98.85$  & $97.88$  & $98.65$  & $97.77$  & $97.62$  &$97.73$   & $97.98$  \\
                       \hline
                       \hline
\multirow{4}{*}{8}   & CD & $0.106$  & $0.141$  & $0.109$  & $0.144$  & $0.142$  & $0.140$  & $0.131$  \\
                              & EE & $9.010$  & $9.047$  & $9.049$  & $9.019$  & $9.058$  & $9.016$  & $8.967$  \\
                              & AEE & $0.753$  & $0.995$  & $0.770$  & $1.012$  & $1.005$  & $0.988$  & $0.920$  \\
                              & AAEE & $0.729$  & $0.978$  & $0.760$  & $0.994$  & $0.986$  & $0.971$  &$0.905$   \\
                              
                       \hline
\multirow{4}{*}{16} & CD & $0.110$  & $0.272$  & $0.127$  & $0.273$  & $0.273$  & $0.259$  & $0.259$  \\
                              & EE & $16.570$  & $18.081$  & $17.883$  & $18.144$  & $18.143$  & $18.017$  & $17.997$  \\
                              & AEE & $0.523$  & $0.962$  & $0.477$  & $0.969$  & $0.968$  & $0.913$  & $0.909$  \\
                              & AAEE & $0.509$  & $0.944$  & $0.470$  & $0.948$  & $0.946$  & $0.897$  & $0.893$  \\
                              
                       \hline
\multirow{4}{*}{32} & CD & $0.090$  & $0.432$  & $0.074$  & $0.534$  & $0.523$  & $0.497$  & $0.482$  \\
                              & EE & $26.934$  & $34.407$  & $32.802$  & $36.344$  & $36.343$  & $36.332$  & $36.332$  \\
                              & AEE & $0.286$  & $0.886$  & $0.187$  & $0.948$  & $0.928$  & $0.881$  & $0.855$  \\
                              & AAEE & $0.283$  & $0.867$  & $0.184$  &$0.927$   &$0.906$   & $0.861$   &$0.838$   \\
                       \hline
\end{tabular}%
}
\vspace{-5pt}
\end{table}

\noindent\textbf{\underline{Adapt PTCs to Different Foundry PDKs}}.~ 
Our method can flexibly adapt different device PDKs. 
We replaced the GF PDK with a customized PDK~\cite{NP_JAP2024_Zhang} in Table~\ref{tab:ComparePDK}.
For a 16$\times$16 PTC size, we applied a new area constraint of [2.208, 15.197] mm$^{2}$. 
The best-performing searched solution, \name-a0, shows 8.26$\times$ higher AAEE than MZI and MMI arrays and 1.04$\times$ compared with Butterfly mesh.
Our solutions use multi-port couplers to enhance information mixing while having fewer blocks and crossings, thus achieving a better balance between expressivity and efficiency.

\begin{table}
\centering
\caption{16$\times$16 PTCs on customized PDKs with MNIST accuracy.
}
\vspace{-10pt}
\label{tab:ComparePDK}
\resizebox{\columnwidth}{!}{%
\begin{tabular}{c|ccc|cccc}
\hline
Metrics & MZI~\cite{NP_NATURE2017_Shen}  & Butterfly~\cite{NP_ACS2022_Feng} & MMI~\cite{NP_SciRep2024_zelaya} & \name-a0       & \name-a1       & \name-a2       & \name-a3         \\ \hline
Area(O+E) &7.51+0.33  &1.25+0.33  &16.90+0.33  &1.19+0.33 &1.22+0.33 &1.21+0.33 &1.71+0.33 \\
                              Power &223.59  &219.93  &219.45  &218.37 &218.54 &218.30 &218.93 \\
                              Latency &100.00  &100.00  &100.00  &100.00 &100.00 &100.00 &100.00 \\
                              \text{Accuracy} &98.74  &98.16  &98.58  &97.67 &98.02 &98.08 &98.18 \\ \hline
                              CD &0.653  &3.242  &0.297  &3.370 &3.301 &3.322 &2.507 \\
                              EE &22.899  &23.280  &23.330 &23.446 &23.429 &23.454 &23.387 \\
                              AEE  &2.919  &14.744  &1.354  &15.434 &15.105 &15.215 &11.450 \\
                              AAEE  &2.882  &14.473  &1.336  &15.074 &14.806 &14.923 &11.242 \\
                              \hline
\end{tabular}%
}
\vspace{-10pt}
\end{table}

\noindent\textbf{\underline{Generalizability to New ONNs and Datasets}}.~
It is important that our searched topology can be \textbf{generalized} to new ONNs and datasets other than the one used for the search.
We train our searched PTC structures on various new benchmarks in Table~\ref{tab:CompareDataset}. 
Though \name-a0 is searched on 2-layer CNN and MNIST, it \emph{maintains superior performance and efficiency} on more complicated models and datasets, showing an average of 1.6$\times$ higher AAEE than manual baselines.

\begin{table}[]
\centering
\caption{16$\times$16 \name-a0 is searched on CNN-MNIST and adapted to new benchmarks with GF PDKs. AAEE is shown.
}
\vspace{-10pt}
\label{tab:CompareDataset}
\resizebox{0.85\columnwidth}{!}{%
\begin{tabular}{cc|ccc|c}
\toprule
Model                    & \multicolumn{1}{c|}{Dataset} & MZI~\cite{NP_NATURE2017_Shen} & Butterfly~\cite{NP_ACS2022_Feng} & MMI~\cite{NP_SciRep2024_zelaya}& \name-a0 \\ \midrule
\multirow{1}{*}{CNN}   & \multicolumn{1}{c|}{FMNIST}   & 0.472   & 0.852  &0.432     &0.853     \\ 
\midrule
\multirow{1}{*}{VGG8} & \multicolumn{1}{c|}{CIFAR10}   & 0.429   & 0.744   &0.382     &0.769     
    \\ 
\midrule
\multirow{1}{*}{ResNet20}   & \multicolumn{1}{c|}{SVHN}   & 0.491   &0.884    &0.445     & 0.890    \\ 
\bottomrule
\end{tabular}%
}
\vspace{-3pt}
\end{table}

\section{Conclusion}
\label{sec:Conclusion}
In this work, we propose a zero-shot multi-objective evolutionary circuit topology search framework \name to explore Pareto-optimal photonic tensor core designs.
In an augmented design space with multi-port couplers, our customized evolutionary algorithm simultaneously optimizes accuracy, compute density, and efficiency, honoring various hardware constraints with balanced exploration and exploitation.
By paying less than a 3-hour search cost, our method can obtain tens of diverse Pareto-optimal circuit topologies, outperforming state-of-the-art manual designs with 2$\times$ higher accuracy weighted area-energy efficiency, with great flexibility and generalizability to more complicated applications and new hardware specifications.


\begin{thebibliography}{10}

\bibitem{NP_NATURE2017_Shen}
Yichen Shen, Nicholas~C. Harris, Scott Skirlo, et~al.
\newblock Deep learning with coherent nanophotonic circuits.
\newblock {\em Nature Photonics}, 2017.

\bibitem{NP_PIEEE2020_Cheng}
Q.~{Cheng}, J.~{Kwon}, M.~{Glick}, M.~{Bahadori}, L.~P. {Carloni}, and K.~{Bergman}.
\newblock {Silicon Photonics Codesign for Deep Learning}.
\newblock {\em Proceedings of the IEEE}, 2020.

\bibitem{NP_NaturePhotonics2021_Shastri}
Bhavin~J. Shastri, Alexander~N. Tait, T.~Ferreira de~Lima, Wolfram H.~P. Pernice, Harish Bhaskaran, C.~D. Wright, and Paul~R. Prucnal.
\newblock {Photonics for Artificial Intelligence and Neuromorphic Computing}.
\newblock {\em Nature Photonics}, 2021.

\bibitem{NP_ACS2022_Feng}
Chenghao Feng, Jiaqi Gu, Hanqing Zhu, Zhoufeng Ying, Zheng Zhao, et~al.
\newblock A compact butterfly-style silicon photonic--electronic neural chip for hardware-efficient deep learning.
\newblock {\em ACS Photonics}, 9(12):3906--3916, 2022.

\bibitem{NP_Science2024_Xu}
Zhihao Xu, Tiankuang Zhou, Muzhou Ma, ChenChen Deng, Qionghai Dai, and Lu~Fang.
\newblock Large-scale photonic chiplet taichi empowers 160-tops/w artificial general intelligence.
\newblock {\em Science}, 384(6692):202--209, 2024.

\bibitem{NP_SciRep2017_Tait}
Alexander~N. Tait, Thomas~Ferreira de~Lima, Ellen Zhou, et~al.
\newblock Neuromorphic photonic networks using silicon photonic weight banks.
\newblock {\em Sci. Rep.}, 2017.

\bibitem{NP_Nature2021_Xu}
Xingyuan Xu, Mengxi Tan, Bill Corcoran, Jiayang Wu, Andreas Boes, Thach~G. Nguyen, Sai~T. Chu, Brent~E. Little, Damien~G. Hicks, Roberto Morandotti, Arnan Mitchell, and David~J. Moss.
\newblock {11 TOPS photonic convolutional accelerator for optical neural networks}.
\newblock {\em Nature}, 2021.

\bibitem{NP_Nature2021_Feldmann}
Johannes Feldmann, Nathan Youngblood, Maxim Karpov, Helge Gehring, Xuan Li, Maik Stappers, Manuel~Le Gallo, Xin Fu, Anton Lukashchuk, Arslan Raja, Junqiu Liu, David Wright, Abu Sebastian, Tobias Kippenberg, Wolfram Pernice, and Harish Bhaskaran.
\newblock Parallel convolutional processing using an integrated photonic tensor core.
\newblock {\em Nature}, 2021.

\bibitem{NP_NatureComm2022_Zhu}
H.H. Zhu, J.~Zou, H.~Zhang, et~al.
\newblock Space-efficient optical computing with an integrated chip diffractive neural network.
\newblock {\em Nature Commun.}, 2022.

\bibitem{NP_SciRep2024_zelaya}
Kevin Zelaya, Matthew Markowitz, and Mohammad-Ali Miri.
\newblock The goldilocks principle of learning unitaries by interlacing fixed operators with programmable phase shifters on a photonic chip.
\newblock {\em Scientific Reports}, 14(1):10950, 2024.

\bibitem{NP_NatureComm2021_Zhang}
H.~Zhang, M.~Gu, X.~D. Jiang, J.~Thompson, H.~Cai, S.~Paesani, R.~Santagati, A.~Laing, Y.~Zhang, M.~H. Yung, Y.~Z. Shi, F.~K. Muhammad, G.~Q. Lo, X.~S. Luo, B.~Dong, D.~L. Kwong, L.~C. Kwek, and A.~Q. Liu.
\newblock {An optical neural chip for implementing complex- valued neural network}.
\newblock {\em Nature Communications}, 2021.

\bibitem{NP_APLML2024_Gu}
Jiaqi Gu, Hanqing Zhu, Chenghao Feng, Zixuan Jiang, Ray~T. Chen, and David~Z. Pan.
\newblock M3icro: Machine learning-enabled compact photonic tensor core based on programmable multi-operand multimode interference.
\newblock {\em APL Machine Learning}, 2024.

\bibitem{NP_ASPDAC2020_Gu}
Jiaqi Gu, Zheng Zhao, Chenghao Feng, et~al.
\newblock Towards area-efficient optical neural networks: an \text{FFT}-based architecture.
\newblock In {\em Proc.~ASPDAC}, 2020.

\bibitem{NP_TCAD2020_Gu}
Jiaqi Gu, Zheng Zhao, Chenghao Feng, et~al.
\newblock {Towards Hardware-Efficient Optical Neural Networks: Beyond FFT Architecture via Joint Learnability}.
\newblock {\em IEEE TCAD}, 2020.

\bibitem{NP_DAC2022_Gu}
Jiaqi Gu, Hanqing Zhu, Chenghao Feng, Zixuan Jiang, Mingjie Liu, Shuhan Zhang, Ray~T. Chen, and David~Z. Pan.
\newblock {ADEPT: Automatic Differentiable DEsign of Photonic Tensor Cores }.
\newblock In {\em Proc.~DAC}, 2022.

\bibitem{NP_Jlt2012_Zhou}
Junhe Zhou and Philippe Gallion.
\newblock Operation principles for optical switches based on two multimode interference couplers.
\newblock {\em IEEE Journal of Lightwave Technology}, 30(1), January 2012.

\bibitem{NAS_ICLR2023_Li}
Guihong Li, Yuedong Yang, Kartikeya Bhardwaj, and Radu Marculescu.
\newblock Zico: Zero-shot nas via inverse coefficient of variation on gradients.
\newblock {\em arXiv preprint arXiv:2301.11300}, 2023.

\bibitem{NN_ICCV2021_Lin}
Ming Lin, Pichao Wang, Zhenhong Sun, Hesen Chen, Xiuyu Sun, Qi~Qian, Hao Li, and Rong Jin.
\newblock Zen-nas: A zero-shot nas for high-performance deep image recognition.
\newblock In {\em Proc.~ICCV}, 2021.

\bibitem{NAS_ICLR2021_abdelfattah}
Mohamed~S Abdelfattah, Abhinav Mehrotra, {\L}ukasz Dudziak, and Nicholas~D Lane.
\newblock Zero-cost proxies for lightweight nas.
\newblock {\em arXiv preprint arXiv:2101.08134}, 2021.

\bibitem{NP_HPCA2024_Zhu}
Hanqing Zhu, Jiaqi Gu, Hanrui Wang, Zixuan Jiang, Zhekai Zhang, Rongxin Tang, Chenghao Feng, Song Han, et~al.
\newblock Lightening-transformer: A dynamically-operated photonic tensor core for energy-efficient transformer accelerator.
\newblock In {\em Proc.~HPCA}, 2024.

\bibitem{NN_TEC2002_Deb}
K.~Deb, A.~Pratap, S.~Agarwal, and T.~Meyarivan.
\newblock A fast and elitist multiobjective genetic algorithm: Nsga-ii.
\newblock {\em IEEE Transactions on Evolutionary Computation}, 6(2):182--197, 2002.

\bibitem{NN_MNIST1998}
Y.~{LeCun}.
\newblock The {MNIST} database of handwritten digits.
\newblock \url{http://yann.lecun.com/ exdb/mnist/}, 1998.

\bibitem{NN_FashionMNIST2017}
Han Xiao, Kashif Rasul, and Roland Vollgraf.
\newblock {Fashion-MNIST: a Novel Image Dataset for Benchmarking Machine Learning Algorithms}.
\newblock {\em Arxiv}, 2017.

\bibitem{NN_svhn2011}
Yuval Netzer, Tao Wang, Adam Coates, Alessandro Bissacco, et~al.
\newblock {Reading Digits in Natural Images with Unsupervised Feature Learning}.
\newblock In {\em Proc.~NIPS}, 2011.

\bibitem{NN_cifar2009}
Alex Krizhevsky, Geoffrey Hinton, et~al.
\newblock Learning multiple layers of features from tiny images.
\newblock 2009.

\bibitem{NP_OFC2020_Rakowski}
Michal Rakowski, Colleen Meagher, Karen Nummy, Abdelsalam Aboketaf, Javier Ayala, Yusheng Bian, Brendan Harris, Kate Mclean, Kevin McStay, Asli Sahin, Louis Medina, Bo~Peng, Zoey Sowinski, Andy Stricker, Thomas Houghton, Crystal Hedges, Ken Giewont, Ajey Jacob, Ted Letavic, Dave Riggs, Anthony Yu, and John Pellerin.
\newblock 45nm cmos — silicon photonics monolithic technology (45clo) for next-generation, low power and high speed optical interconnects.
\newblock In {\em 2020 Optical Fiber Communications Conference and Exhibition (OFC)}, pages 1--3, 2020.

\bibitem{NP_JAP2024_Zhang}
Meng Zhang, Dennis Yin, Nicholas Gangi, et~al.
\newblock Tempo: efficient time-multiplexed dynamic photonic tensor core for edge ai with compact slow-light electro-optic modulator.
\newblock {\em Journal of Applied Physics}, 135(22), 2024.

\end{thebibliography}

\end{document}